\documentclass{aa}
\usepackage{natbib,epsfig,graphicx,ulem}
\usepackage{color,soul}
\bibpunct{(}{)}{;}{a}{}{,}
\def\mathnew{\mathsurround=0pt}
\def\simov#1#2{\lower .5pt\vbox{\baselineskip0pt \lineskip-.5pt
\ialign{$\mathnew#1\hfil##\hfil$\crcr#2\crcr\sim\crcr}}}
\def\simgreat{\mathrel{\mathpalette\simov >}}

\def\arcmin{\hbox{$^\prime$}}
\def\arcsec{\hbox{$^{\prime\prime}$}}

\def\MeV{Me\kern-0.11em V}
\def\keV{ke\kern-0.11em V}
\def\arcsec{\hbox{$^{\prime\prime}$}}
\def\cha{{\it Chandra\/}}
\def\xmm{{\it XMM-Newton\/}}

\def\spi{{\it Spitzer\/}}

\def\Ms{M$_{\odot}$\/}
\def\ecmss{ergs cm$^{-2}$ s$^{-1}$}
\def\es{ergs s$^{-1}$}

\begin{document}

\title{Cluster and cluster galaxy evolution history from IR to X-ray
observations of the young cluster RX J1257.2+4738 at z=0.866
\thanks {Partly based on observations with \cha\ which is operated by
the Harvard-Smithsonian Center for Astrophysics under contract with
NASA. Partly based on observations obtained at the Gemini Observatory
which is operated by the Association of Universities for Research in
Astronomy, Inc., under a cooperative agreement with the NSF on behalf of
the Gemini partnership: the National Science Foundation (United States),
the Science and Technology Facilities Council (United Kingdom), the
National Research Council (Canada), CONICYT (Chile), the Australian
Research Council (Australia), Ministerio da Ciencia e Tecnologia
(Brazil) and SECYT (Argentina). Partly based on data collected at the
Subaru Telescope, which is operated by the National Astronomical
Observatory of Japan. Partly based on observations obtained with the
Apache Point Observatory 3.5-meter telescope, which is owned and
operated by the Astrophysical Research Consortium (ARC). Partly based on
observations made with the \spi\ Space Telescope, which is operated by
the Jet Propulsion Laboratory, California Institute of Technology under
a contract with NASA. Partly based on observations obtained with {\it
XMM-Newton}, an ESA science mission with instruments and contributions
directly funded by ESA Member States and the USA (NASA). Partly based on
observations obtained at the Observatoire de Haute Provence (CNRS) with
the 1.93-meter telescope and the CARELEC instrument.}}

\author {M. P. Ulmer\inst{1,2} \and 
C. Adami\inst{1} \and
G.B. Lima Neto\inst{3} \and
F. Durret\inst{4} \and
G. Covone\inst{5} \and
O. Ilbert\inst{1,6} \and
E.S. Cypriano\inst{7,3} \and
S.S.~Allam\inst{8} \and
R.G.~Kron\inst{9} \and
W.A.~Mahoney\inst{10} \and
R. Gavazzi\inst{4} 
}

\offprints{C. Adami \email{christophe.adami@oamp.fr}}

\institute{
LAM, P\^ole de l'Etoile Site Ch\^ateau-Gombert, 38 rue Fr\'ed\'eric
Juliot-Curie, 13388 Marseille Cedex 13, France
\and
Department of Physics and Astronomy, Northwestern University,
2131 Sheridan Road, Evanston IL 60208-2900, USA
\and 
Instituto de Astronomia, Geof\'{\i}sica e C. Atmosf./USP, R. do Mat\~ao 1226, 
05508-090 S\~ao Paulo/SP, Brazil
\and
Institut d'Astrophysique de Paris, CNRS, UMR~7095, Universit\'e Pierre
et Marie Curie, 98bis Bd Arago, 75014 Paris, France
\and 
Universit\`{a} di Napoli `` Federico II'', Dipartimento di Scienze Fisiche and 
INAF -- Observatorio Astronomico di Capodimonte, v. Moiariello 16, 80131 Napoli, Italy
\and
Institute for Astronomy, 2680 Woodlawn Dr., University of Hawaii, Honolulu, 
Hawaii 96822, USA
\and
Department of Physics and Astronomy, University College London,Gower
Street, London WC1E 6BT, UK
\and
Fermi National Accelerator Laboratory, MS 127, P.O. Box 500, Batavia, 
IL 60510, USA
\and
University of Chicago, Department of Astronomy and Astrophysics, 5640 South
Ellis Avenue, Chicago, IL 60637, USA
\and 
California Institute of Technology, Spitzer Science Center, MS 314-6, 1200 East California Blvd,
Pasadena , CA 91125, USA
}

\date{Accepted , Received ; Draft printed: \today}

\authorrunning{Ulmer et al.}

\titlerunning{The X-ray emitting distant cluster RX J1257.2+4738}

\abstract
{The cosmic time around the z$\sim$1 redshift range appears crucial in
the cluster and galaxy evolution, since it is probably the epoch of the
first mature galaxy clusters.  Our knowledge of the properties of the
galaxy populations in these clusters is limited because only a handful
of z$\sim$1 clusters are presently known.}
{In this framework, we report the discovery of a z$\sim$0.87 cluster
and study its properties at various wavelengths.}
{We gathered X-ray and optical data (imaging and spectroscopy), and near
and far infrared data (imaging) in order to confirm the cluster nature
of our candidate, to determine its dynamical state, and to give insight
on its galaxy population evolution.}
{Our candidate structure appears to be a massive z$\sim$0.87 dynamically
young cluster with an atypically high X-ray temperature as compared to
its X-ray luminosity. It exhibits a significant percentage ($\sim 90$\%)
of galaxies that are also detected in the 24$\mu$m band.}
{The cluster RXJ1257.2+4738 appears to be still in the process of
collapsing. Its relatively high temperature is probably the consequence
of significant energy input into the intracluster medium besides the
regular gravitational infall contribution. A significant part of its
galaxies are red objects that are probably dusty with on-going star
formation.}


\maketitle

\section{Introduction}\label{sec:intro}

It is becoming increasingly clear that the redshift range between
$\sim$0.8 and 1.0 is particularly interesting for comparing star
formation histories of galaxies in clusters and in the field, as well as
studying the relationship of galaxy infall and heating of the intracluster
medium (ICM; e.g. Marcillac et al. 2007, Gilbank et al. 2008, Kocevski et al. 2009).

On the observational side, we might expect selection effects to favor
finding the most X-ray luminous 
clusters when searching for clusters at high redshift.  This is not the
case, however.  For example, the compilation by Ota et al. (2006) shows
that 50\% of the cluster sample between $z$ = 0.3 and 0.56 have a
bolometric X-ray luminosity (L$_{\rm{X,bol}}$) greater than 10$^{45}$
ergs s$^{-1}$ compared to only 20\% of the sample above $z = 1$
(Fassbender 2008, Kocevski et al. 2009).  These X-ray luminosities
suggest that beyond z=1, clusters are growing, but have not reached
their peak X-ray luminosity.

In comparison, we {\it estimate} the free fall times for clusters (with a
typical radius of 1 Mpc and masses between $(0.5- 5)\times 10^{14}$
M$_{\odot}$) of the order of 1$-$2 Gyr (Sarazin 1986). 
The elapsed time between $z = 2$ 
and $z = 1$ is about 2.6 Gyrs.  Thus, at redshift range
near $\sim 1$ we can expect mainly to find clusters of $(0.5- 5) \times
10^{14}$ M$_{\odot}$ that have just completed their initial infall
(see Sadeh $\&$ Rephaeli 2008 and references therein).

Since X-ray detectable clusters near $z = 1 $ are probably just
forming large enough systems to become detectable, a study of the
relationship between the galaxy population and the X-ray emission is
particularly instructive for refining models of both large scale
structure evolution and galaxy evolution. See Ettori et al. 2004b,
Andreon 2006, Ota et al. 2006, for comprehensive studies of X-ray
detected clusters and cluster evolution, and Cooper et al. 2008, Yan et
al. 2008 for recent studies on galaxy formation and evolution related to
clusters of galaxies at z$\sim 0.8-0.9$.

In the process of our search for moderately distant (i.e., $z \sim 1$)
clusters of galaxies that are detectable via their X-ray emission (Adami
et al. 2000, Romer et al. 2000, Ulmer et al. 2005, Adami et al. 2007),
we report here the discovery of a cluster with $z = 0.866$, RX
J1257.2+4738 (hereafter for brevity referred to as RX J1257).  The
cluster was first detected in X-rays in the {\it ROSAT} data archive,
and the i$'$-band and Ks-band follow-up found a concentration of red
galaxies that could be explained by the existence of a cluster with $z
\simgreat 0.6$.

This cluster is particularly interesting because it contains several red
galaxies with [OII] 3727 emission as well as several others that are
detected both in the \spi\ MIPS 24 $\mu$m and \spi\ IRAC bands.  Without
the detected emission lines or the MIPS emission, these red galaxies
might have been characterized as early type galaxies, or red-dead
galaxies (Elbaz et al. 2007). The emission lines imply young galaxies
undergoing star formation and the high MIPS flux implies dusty late type
galaxies with ongoing star formation.  We report here a detailed study
based on extensive 
observations including \cha, \xmm, {\it Spitzer}, Gemini, Subaru, and
ARC.  Also, as a by-product, we found candidates for lensed (and hence
magnified) background galaxies or AGNs that are at least at $z > 3$, and
could be at redshifts as high as 10.  These will be presented in a
separate later paper. A preliminary version of the results reported here
were given by Ulmer et al (2008).

Throughout, we assume a concordant $\Lambda$CDM cosmology with H$_0$=71 km
s$^{-1}$ Mpc$^{-1}$, $\Omega_{\Lambda,0}$ = 0.73, $\Omega_{m,0} = 0.27$,
from which we compute the angular scale as 7.72 kpc/arcsec, a luminosity
distance (D$_{\rm L}$) = 5544.7 Mpc and an angular distance (D$_{\rm
A}$) = 1592.4 Mpc (Wright 2006).

\section{Observations, Analysis and Results}
\subsection{X-rays}

The net \cha\ exposure time was 37 ksec in very faint mode with
ACIS-S3 (back illuminated chip) on July 24, 2007.

The \xmm\  net exposure time, after high background filtering was
16 ksec (11 and 5 ksec for each usable exposure). The nominal
exposure time (83 ksec allocated) was heavily polluted by flares. The
observations were made with the Thin 1 filter in Prime full window on May
24th through 26th, 2007.

\subsubsection{Chandra data processing}

The data were reduced with the CIAO version
3.0.1\footnote{\texttt{http://asc.harvard.edu/ciao/}} following the Standard
Data Processing, producing new level 1 and 2 event files.

We further filtered the level 2 event file, keeping only \textit{ASCA} grades
0, 2, 3, 4 and 6, and restricted our data reduction and analysis to the
back-illuminated chip, ACIS-S3. We checked that no afterglow was present and
applied the Good Time Intervals (GTI) supplied by the pipeline. We then
checked for flares in the light-curve in the [10--12 keV] band. No flares were
detected.


\subsubsection{\xmm\ data processing}
The \xmm\ data were so badly contaminated with flare events, that in
addition to the standard processing tools,
visual inspection of the overall count rate was needed as well in order
to select only good data. 

RX J1257 was observed in standard Full Frame mode with the ``thin''
filter with the two EPIC MOS1 and MOS2 and the PN detectors. The basic
data processing (the ``pipeline'' removal of bad pixels, electronic
noise and correction for charge transfer losses) was done with package
SAS V5.3, thus creating calibrated event files for each detector.

For the MOS1 and MOS2 cameras, following the standard procedure, we have
discarded the events with FLAG~$\neq 0$ and PATTERN~$> 12$; for the PN, we have
restricted the analysis to the events with PATTERN $\le$ 4 and Flag =0.

With the cleaned event files, we have created the redistribution matrix file
(RMF) and ancillary response file (ARF) with the SAS tasks \texttt{rmfgen} and
\texttt{arfgen} for each camera and for each region that we have analyzed.


\subsubsection{Imaging}
\label{Imaging}

We detect the diffuse emission of the cluster both in \cha\ and \xmm\
observations. If we model the \xmm\ emission with an ellipse it has a semi
major axis of 1.2~arcmin and a semi-minor axis of 1.0~arcmin (556 kpc and 463
kpc at $z = 0.866$). The major axis runs along the E-W direction and the
cluster is centered at 12h 57m 12.2s, +47$^{\circ}$ 38$'$ 06.5$''$ (J2000).
A point source we detected with \cha\ appears embedded inside the
diffuse emission. We will discuss this source in $\S$~\ref{XRS} and in
$\S$~\ref{pt_src} below.

The X-ray emission is statically significant. The count rate of the \cha~
observation, inside the ellipse where the cluster spectrum is extracted (see
below), is $0.0065 \pm 0.0002$~cnt/s/arcmin$^{2}$ (where the point sources are
already masked out). On the same CCD, we estimate the background count rate as
$0.00495 \pm 0.00015$~cnt/s/arcmin$^{2}$. Therefore, the \textit{net} count
rate of the cluster in the central $1.2' \times 1'$ region is $0.0016 \pm
0.0004$~cnt/s/arcmin$^{2}$, or $4\sigma$ above zero. Combining all XMM
observations (all detectors and exposures), we obtain for the same region
(masking the point sources taking into account the larger XMM PSF) a
\textit{net} count rate of $0.0059 \pm 0.0002$~cnt/s/arcmin$^{2}$, well above
a null count rate.

We are dealing here with a densely populated field of active objects,
which dominates the X-ray emission of the area. We therefore
\textit{masked} these point source contributions to determine the diffuse emission of
the cluster. The ability to detect and verify the point-like nature of these
objects was
impossible to do using only \xmm\ data due to a lack of angular resolution.
We  therefore used the \cha\ data.
The point source masks removed over 99\% of the encircled point source flux and thus the
remaining flux was point source contaminated at a level well below
statistical uncertainty in our derived best fit spectral shapes and
overall luminosity.

For all the relevant masked X-ray point sources, we found optical counterparts
that were bright enough to apply {\it LePhare}\footnote{S. Arnouts \& O. Ilbert; ~\\
www.oamp.fr/people/arnouts/LE\_PHARE.html}.  
For those objects classified as galaxies, we estimated the probability 
that they
belong to the cluster based on
the probability distribution function of the photometric redshifts
(hereafter the PDF). We determined the PDFs with the {\it LePhare
software} The PDF gives the
probability for an object to be at a given redshift. The probability for
an object to be in a redshift interval is therefore obtained by
integration over this interval and normalization to the total
probability. We give below the list of the prominent X-ray point
sources, the probability that they are in the cluster redshift interval,
and the LePhare spectral classification.

The main point sources showing up both in \cha\ and \xmm\ data are:

\noindent -- ($\alpha$=194.2680, $\delta$=47.6377) associated
with a bright galaxy. The probability of
this bright X-ray object to be in the [0.82,0.90] redshift interval (potentially 
a cluster member) is lower than 0.2\%. This is
probably a late type active galaxy (type 19 from the 
$LePhare$ spectral data base) equivalent to a starburst; 

\noindent -- ($\alpha$=194.2930, $\delta$=47.6392) associated
with a galaxy. The probability of
this object to be in the [0.82,0.90] redshift interval is lower than 0.4\%. 
This is probably a very active object, $LePhare$ gives a better $\chi ^2$ 
with a QSO template than with a galaxy template; 

\noindent -- ($\alpha$=194.2930, $\delta$=47.6339) associated
with a {\it blend} of 3 objects within a 2\arcsec\ long region. 
This object is discussed in detail in \S  3.3.; 

\noindent -- ($\alpha$=194.2570, $\delta$=47.6251) associated
with an active galaxy at z=0.662 from our Gemini spectroscopic
catalog. This galaxy exhibits [OII] and strong [OIII] lines, as well as 
broad H$\gamma$ and H$\beta$ emissions, characteristic of an active object; 

\noindent -- ($\alpha$=194.2950, $\delta$=47.6476) which is probably a
local star (not in the Simbad data base), since $LePhare$ gives the best
$\chi ^2$ with a star template.  Using the probability
distribution function given by a galaxy template, the probability that
the object's redshift is in the [0.82,0.90] redshift interval is lower
than 9\%;

\noindent -- ($\alpha$=194.3051, $\delta$=47.6132) associated with the
GMOS guiding object (i$'$ SDSS mag = 12.39). We observed its spectrum at
the Observatoire de Haute Provence 1.93m telescope with the CARELEC
spectrograph. The spectrum is typical of a local old star;

\noindent -- ($\alpha$=194.2914, $\delta$=47.6618) associated with a faint
probably high redshift object (from its PDF)
which only has a probability of less than 1.5\% to be part of the
cluster;

\noindent -- ($\alpha$=194.2876, $\delta$=47.6641) associated with a faint
probably high redshift object (from its PDF)
which only has a probability of less than 0.1\% of being a cluster
member;

\noindent -- ($\alpha$=194.3070, $\delta$=47.6715) associated with a
relatively bright optical object. This object has a non-negligible
probability of being at the cluster redshift (12\%) but since it is
located at a projected distance of over 1 Mpc from the cluster center,
it is probably not directly related with the structure itself. We cannot exclude,
however, that the galaxy is undergoing cluster infall.

\subsubsection{X-ray spectroscopy of the diffuse source}
\label{XRS}

The background spectra were extracted from the same observations, on the
same CCD chip, from a region close to the cluster, free of point
sources. In this way, we took into account any
count rate anomaly in the background.  The center of the background
region is only about 3.5~arcmin from the source, making the vignetting
effect (for the cosmic X-ray background term) negligible.  The
redistribution and ancillary files (RMF and ARF) were created with the
SAS tasks rmfgen and arfgen for each exposure, taking into account the
extended nature of the source.

We have fit simultaneously spectra from the two PN exposures with the
MEKAL spectrum model of a thermal plasma.  The galactic neutral hydrogen
along the line of sight was taken into account with the ``phabs''
photoelectric absorption model of Balucinska-Church \& McCammon
(1992). The hydrogen column density, N$_{\rm H}$, was fixed to the
galactic value given by the Leiden/Argentine/Bonn (LAB) Survey of
Galactic HI (Kalberla et al. 2005); N$_{\rm
H} = 1.2 \times 10^{20}$ cm$^{-2}$.  We have also fixed the metal
abundance to Z = 0.4Z$_\odot$ and the redshift to z = 0.866 (see the
optical spectroscopy section).  In order to minimize any discrepancy
between \cha\ and \xmm\ at low and high energies, we limited the
spectral fit to the 0.7--7.0 keV band.

\begin{figure}[bht]
\centering
\mbox{\epsfig{file=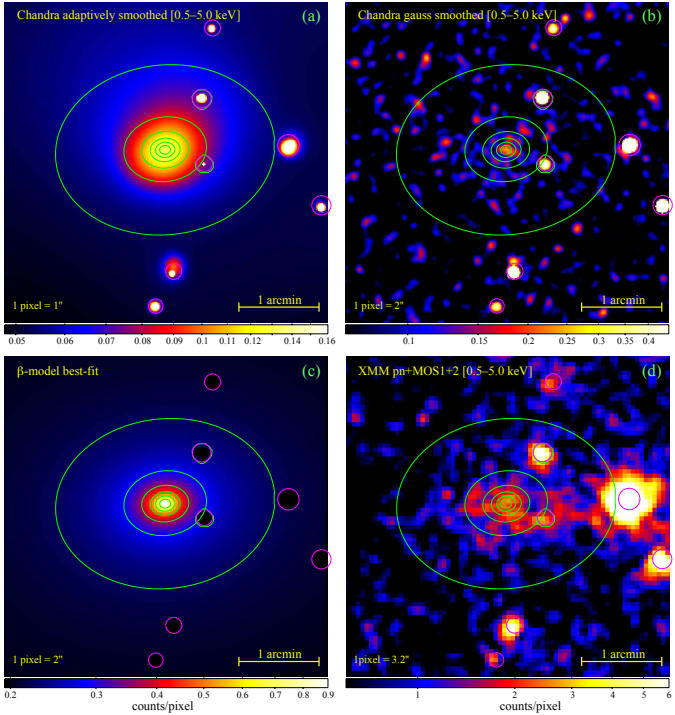,width=3.5in,angle=0}}
\caption{\label{gastao_color} X-ray imaging of Cl~1257 in the
0.5--5.0~keV band. (a)~Adaptively smoothed ACIS-S image with a minimum
signal-to-noise ratio of 3 per smooth beam done with the COAI 3.4
csmooth task; (b)~ACIS-S image smoothed with a fixed width Gaussian
(kernel radius of $10$~arcsec); (c)~best-fit 2D $\beta$-model;
(d)~composite image with all available XMM data smoothed with a Gaussian
kernel (radius $6.4$~arcsec). The green ellipses correspond to
logarithmically spaced isocontours of the best-fit $\beta$-model.  The
magenta circles show the masked regions (omitted both in the 2D-fit and
in the spectral fits) corresponding to point-sources detected with \cha\
data. The color bars indicate the imaging scaling in counts per pixel.}
\end{figure}

\begin{figure}[bht]
\centering
\mbox{\epsfig{file=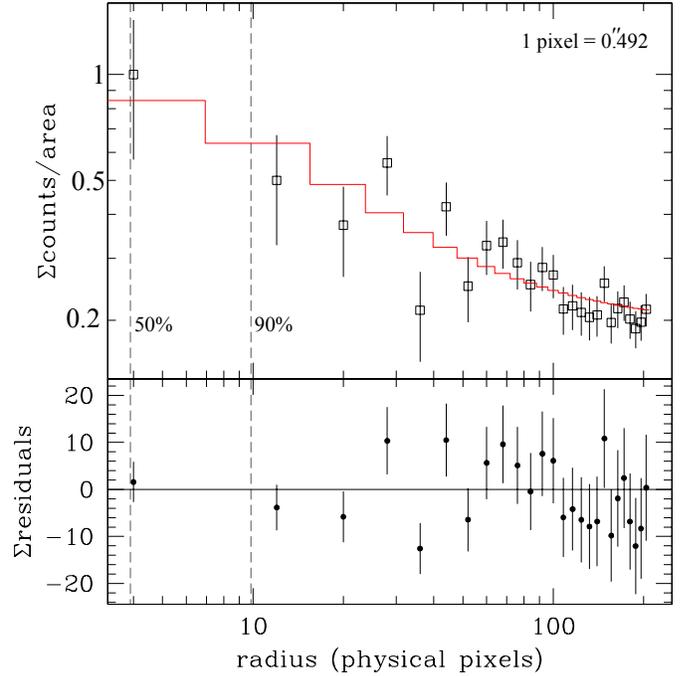,width=3.5in,angle=0}}
\caption{\label{beta} Radial surface brightness profile best fit $\beta$-model to
\cha\ data after point source masking. The vertical dashed lines
indicate the radii enclosing 50\% and 90\% of the energy from
a point source, spread by the \cha\ PSF.}
\end{figure}

Since the X-ray field in the direction of the cluster is heavily contaminated
by X-ray point sources, it is useful to archive the
masking out of these sources as shown in  Fig. \ref{gastao_color}. We show the
resulting best fit $\beta$\ model in Fig. \ref{beta}  based on the
assumption that the cluster has an intrinsically smooth elliptical
surface brightness distribution. We also show in Fig. \ref{beta} the 50\% and
90\% encircled energy radii to demonstrate that the X-ray emission is clearly diffuse.
{\it The assumption of a symmetric $\beta$ model is only necessary
for extracting the spectrum and was not used to determine the true shape
of the X-ray emission}. A detailed discussion of the nature of the
inferences based on the shape of the emission 
is given below. We conclude, though, that we have have indeed detected diffuse
emission. The results of the spectral fitting are given in Table \ref{X_rays}.

\begin{table}[hbt]
\begin{center}
\caption{Summary of \xmm\  Spectral Analysis \label{X_rays}}
\begin{tabular}{|c|c|c|c|}
\hline
kT       &  f$_{\rm{x} }$ (0.5-2.0 keV)  &  L$_{\rm{X,bol} }$  & $\chi^2$/dof$^{\dagger}$ \\
(keV)    & ($10^{-14}\,$erg~s$^{-1}\,$cm$^{-2}$) & ($\rm{h}_{70}^{2} 10^{44}\,$erg s$^{-1}$) & \\
\hline
 & & & \\[-6pt]
 3.6$^{+2.9}_{-1.2}$ &$ 2.2 \pm1.1$ & 2.0 $^{+2.9}_{-1.2}$ & 64.06/64\\[5pt]
\hline
\end{tabular}
~\\ $^{\dagger}$ Best fit thermal bremsstrahlung model; $Z = 0.4 Z_{\odot}$
fixed.
\end{center}
\end{table}

For future reference in the text, we use the spectral fit and 
the formula from Hilton et al. (2007) that is based on
the temperature of the ICM, to derive R$_{\rm v}$ =1.05~Mpc or 2.3~arcmin,
where R$_{\rm v}$ is the virial radius of the cluster.

Given the relatively low number of photons included in the spectrum
($\sim$1250 counts including background), the spectrum alone does not
allow us to discriminate at a statistically significant level between
a thermal and power law spectrum that might be expected from an AGN.

\subsection{Visible, NIR and IR data}

 In order to determine the nature of the galaxies in the cluster as well as to
 measure the redshift of the cluster, we obtained visible, near IR and IR
 observations. We summarize the observations and the results of those 
 observations below.

\subsubsection{{\it Spitzer} data reduction}

\spi\ (Werner et al. 2004) images obtained with the Infrared Array
Camera (IRAC Fazio et al. 2004) were obtained on 2006 Dec
29. The observation consisted of 100-second integrations and 20
dithers giving a net exposure of 2000 seconds in each of the four
channels centered at 3.6, 4.5, 5.8, and 8 $\mu$m and covering a
$\sim 5.\arcmin2 \times 5.\arcmin2$ field. A deep image at 24 $\mu$m was made
on 2007 June 02 with the Multiband Imaging Photometer for
\spi\ (MIPS, Rieke et al. 2004) using 30-second integrations
with a total exposure time of 9,050
seconds. The calibrated mosaic images from the standard
\spi\ pipeline showed satisfactory removal of instrumental
artifacts and were sufficiently clean for the IRAC data analysis.
However, the MIPS 24 $\mu$m pipeline mosaic exhibited artifacts and
gradients, which were significantly reduced through a zodiacal
light subtraction, a self-calibrating flat, and an improved overlap
correction algorithm.
Source fluxes were obtained from these mosaic images
with SExtractor (Bertin $\&$ Arnouts 1996) aperture photometry.

The original units of the Spitzer images (MJy/sr) were translated to Jy
for detected objects taking into account the pixel size to which these
images were rebinned (0.15$\times$0.15~arcsec$^2$) 
and the general directions given in
http://ssc.spitzer.caltech.edu/. \footnote{Magnitudes were then computed following
the same directions and translated from the Vega system to the AB system
for the IRAC bands with the numbers listed in
http://spider.ipac.caltech.edu/staff/gillian/cal.html.}

\subsubsection{Gemini spectroscopy and imaging}

\begin{figure}
\centering
\mbox{\epsfig{file=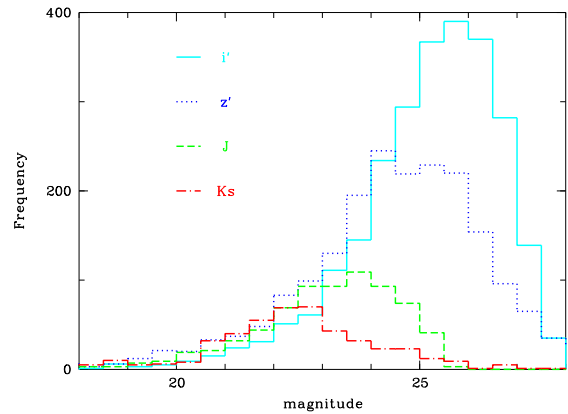,width=2.5in,angle=270}}
\caption{\label{bli} Histograms showing the completeness levels for the i$'$, 
z$'$, J, and Ks bands. }
\end{figure}

The measurement of redshifts in a z$\sim$0.9 galaxy structure required
spectroscopic observations of galaxies as faint as i$'\sim 23$. In order
to achieve this goal, we acquired 3 Gemini north GMOS observations
(GN-2005A-Q-9, GN-2006A-Q-4, and GN-2006B-Q-38). We measured 45
redshifts of galaxies with magnitudes 
between $i'=20$ and 22.6 and
exposure times varying between 3 and 4 hours.  At the same time, we took
advantage of the pre-image observing to acquire 2 deep optical
images of this cluster in the i$'$ and z$'$ bands. The completeness
limit of these images was close to i$'\sim$25 and z$'\sim$23.5 as shown
in Fig.~\ref{bli} (together with the completeness histograms in the J
and Ks bands for comparison).

We obtained images in the i$'$ band under photometric
  conditions. In order to assess the quality of these magnitudes, we
  compared our values with the SDSS estimates. SDSS images cover the
  same area with the same filters and are much less deep
  than our data. However, the brightest objects of our field are also 
  detected in SDSS images. Based on 51 objects in common, we computed the mean
  value of the difference between SDSS i$'$ band magnitudes and our
  estimates, and found a good agreement:

i$'_{SDSS}$ - i$'_{GMOS}$ = $-0.05 \pm$ 0.17

Images in the z$'$ band were observed under non-photometric
conditions. We therefore used the SDSS magnitudes to
rescale our own measurements with 45 objects in common. The resulting
dispersion between SDSS and our estimates is $\pm$0.21 magnitude.

\begin{figure}
\centering
\mbox{\epsfig{file=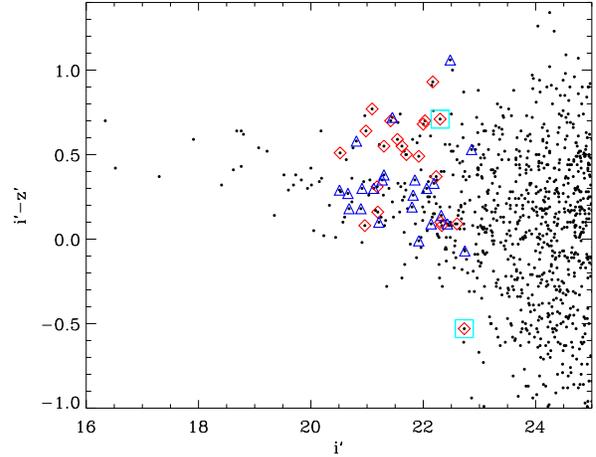,width=2.5in,angle=90}}
\caption{\label{blu} The color magnitude relation computed from i$'$ and
z$'$. The galaxies with spectroscopically determined redshifts that place them in the cluster ($0.850 \leq z \leq 0.874$) are red
diamonds. Galaxies with
 spectroscopically determined  redshifts outside the cluster are blue triangles. The black dots are all 
 galaxies. The green squares mark the only two galaxies at the cluster redshift that were not detected in the MIPS 24 $\mu$m band.}
\end{figure}

We show in Fig. \ref{blu} the color magnitude relation based on the i$'
$ and z$'$ data.  Note that most if not all the scatter in the i$'-$z$'$
direction is attributable to the dispersion in the derived values of
z$'$.  The cluster being at z$\sim$0.87, the 4000~\AA\ break is included
in the i$'$ filter and located just before the z$'$ filter.  We also
superpose on Fig. \ref{blu} the galaxies inside and outside the
z=[0.850,0.874] range. Fig. \ref{blu} shows a possible (given the
dispersion in z$'$) concentration close to i$'$-z$'\sim$0.6 (the red
diamonds in Fig.~\ref{blu}).

We will argue in the following that our cluster is a dynamically young
structure, and thus that many of the galaxies have probably not turned
into ellipticals or S0s. The average color of $i'-z'$=0.6 that we found is
consistent with those of early spirals at $z$ = 0.8 based on Fukugita et
al. (1995) who predicted a color of $\sim$0.6 for these objects.

Fig. \ref{blu} also exhibits a prominent concentration of bright and
probably nearby galaxies with $i'-z'\sim$0.3. We retrieved from the NED
database all the redshifts in a 1 deg$^2$ region centered on the cluster
position. 
The histogram of these redshifts shows two concentrations around
z$\sim$0.03 and z$\sim$0.15. Since these concentrations are present over
the 1 deg$^2$ field of view, we suggest that there are two foreground
galaxy filaments or sheets. In this regard, Fukugita et al. (1995)
predict $i'-z'\sim 0.3$ for early galaxy types at these (both $\sim
0.03$ and $\sim 0.15$) redshifts, which implies these filaments or
sheets are populated by relatively evolved systems.


\subsubsection{Near IR observations}

We also obtained near infrared data (J and Ks bands) in order to fill the gap
between the i$'$ and z$'$ Gemini bands and the Spitzer infrared data.

A J band Subaru MOIRCS image was taken with a 40~min
exposure in 2007. Data were reduced with the standard MOIRCS
pipeline and provided a catalog complete down to J$\sim$23 (see Fig.~\ref{bli}).

We also observed during 2 hours in the Ks band a subfield of the RX J1257 area
at the 3.5m APO telescope with the GRIM camera in 2000. This provided a
catalog complete up to Ks$\sim$22 (see Fig.~\ref{bli}) in a smaller area
(see Fig.~\ref{zphot}).

\subsubsection{Catalog extraction}

\label{catext} 
\begin{figure}
\centering \mbox{\epsfig{file=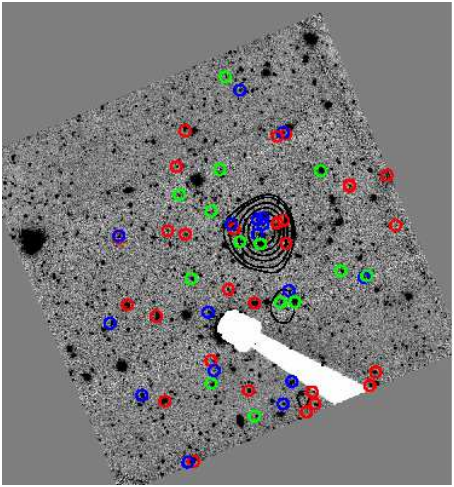,width=6.cm,angle=0}}
\caption{\label{zphot} GMOS $\sim 5.\arcmin2 \times
5.\arcmin2$ field
i$'$-band plus the positions of the galaxies located in the $z$
=[0.82,0.90] interval (derived from {\it LePhare})
over plotted . Black isocontours are computed at the positions of these
galaxies with an adaptive kernel technique and a smoothing window of
7~arcsec$^2$.  The fit templates have types from the $LePhare$ web page
(see footnote in $\S$ \ref{Imaging}), between 0 and 30, 0 being the
earliest type and 30 the latest (star-burst) type. Red, types between
0 and 10, green between 11 and 20, and blue between 21 and 30. }
\end{figure}

The angular resolution of our images ranged from a few
tenths of an arcsec (for the i$'$ GMOS $ \sim 5.\arcmin2 \times 5.\arcmin2$ image) to a few arcsec (for the
IRAC and MIPS images). It is therefore impossible simply to extract object
catalogs independently from all these images and to correlate these
positions because of source confusion. We therefore chose to align and
rebin to the same pixel size (0.15 arcsec) all our images using the
Terapix tools (http://terapix.iap.fr/) and the standard procedure developed
for the CFHTLS fields (McCracken et al. 2008).  We used the
weight maps for each band to produce two areas: one being the common area
of all images and the other being the common area of all images except
the Ks image (which is significantly smaller than all the other
images). This is illustrated in Fig.~\ref{zphot}.

We then extracted an object catalog
from these images with the SExtractor package, in
double-image mode,
with a detection threshold of 1.1 and a minimum number of pixels (above
this threshold) of 3.
This provided an i$'$, z$'$, J, Ks, 3.6$\mu$m, 4.5$\mu$m, 5.8$\mu$m, 8.0$\mu$m, and
24$\mu$m catalog.

\subsubsection{Photometric redshift computations}

We have computed photometric redshifts with the 
$LePhare$ package. 
A full description of this tool is given in Ilbert et al. (2006) and in
the $LePhare$ web page, but here are the salient points.  $LePhare$ is
based on a template-fitting procedure. We used templates from
Polletta et al. (2007) and star-forming templates from
Bruzual $\&$ Charlot (2003). The zero-point calibration has been done on the
basis of the spectroscopic catalog with an iterative procedure described
in section 4.1 of Ilbert et al. (2006), by comparing the predicted
magnitudes from the best fit template and the observed magnitudes.

$LePhare$ required shifts of 0.07 in i$'$, $-$0.24 in z$'$ (data observed
under non-photometric conditions), 0.22 in J, $-$0.44 in Ks (data observed
under non photometric conditions), 0.135 at 3.6~$\mu$m, 0.08 at
4.5~$\mu$m, 0.28 at 5.8~$\mu$m, 0.22 at 8.0~$\mu$m, and 0.0 at 24~$\mu$m.

The relatively high values of these shifts can be explained in part by the
correlation between the shift amplitude and the photometric quality of
the night. 

The shift in the photometry can also be explained in part by the nature of 
the spectral catalog that we used which is 
dominated by galaxies in the RX J1257 cluster.  However,
the spectral shapes of cluster galaxies are not well
characterized in the only available {\it LePhare} template library which
contains only
{\it field galaxy} templates.  
Experience has shown
that comparing field templates to cluster galaxies usually leads to high
shift values. For example, similar shifts of 0.15 were applied to the
Coma cluster photometric data by Adami et al. (2008).

A set of fit parameters was then produced and we are mainly interested
in multicolor type (based on a classification in a 9 magnitude-color
space), redshift, and PDF estimates.
We first assess the quality of the photometrically determined $z$
($z_p$, photo-z) by comparison with the spectroscopically determined $z$
($z_s$), assuming that the uncertainties in $z_s$ are negligible.  If
$\Delta z = z_s - z_p$, we can estimate the redshift accuracy from
$\sigma _{\Delta z/(1+z)}$ using the normalized median absolute
deviation (NMAD Hoaglin et al. 1983) defined as $1.48 \times$
median$(|z_p - z_s|/(1+z_s))$.  This dispersion estimate is robust with
respect to catastrophic errors 
(i.e. objects with $|z_p - z_s|/(1+z_s) > 0.15$).  The percentage of
catastrophic errors is denoted by $\eta$.

\begin{figure}
\centering
\mbox{\epsfig{file=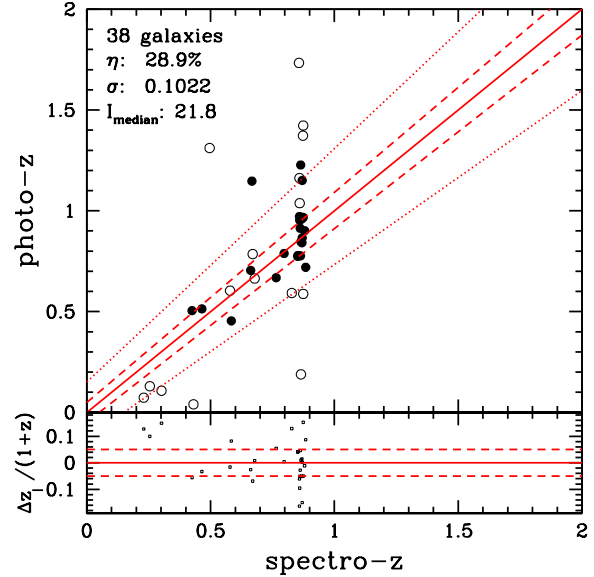,width=3.in,angle=0}}
\caption{\label{zphotcheck} Spectroscopic versus photometric redshifts. Open
  symbols: multi-peak PDFs, where the
  redshift chosen is the one with highest probability; filled symbols: 
 single-peak PDFs.  See text.
 }
\end{figure}

We show in Fig.~\ref{zphotcheck} the relation between the estimated
photometric redshift and the spectroscopic value of the redshift.
Beyond the general agreement, we see that the photometric
redshift of several galaxies is not very well determined. Most of these
galaxies have several peaks in their PDFs (open symbols in
Fig.~\ref{zphotcheck}) and this shows the need to consider the PDF instead of
the best fit value of the photometric redshift.  I$_{\rm{median}}$ is
the median magnitude of the spectroscopic sample taken as reference.  We
used 1/(1+z) weighting of the residuals to be consistent with previous
work such as Ilbert et al. (2006).

Based on the PDF, we computed the integrated probability for each galaxy
to have a redshift inside a given interval. From the spectroscopic
catalog, we determined that the optimal interval is z=[0.82,0.90]. When
predicting that a galaxy is within this interval on the basis of the
PDF, our success rate is close to 80\%. When predicting that a galaxy
is outside this interval, our success rate is close to 65\%.

We therefore extracted a candidate list of galaxies with a high
probability of being part of the cluster by being included in the
interval $z$ = [0.82,0.90]. We plot in Fig.~\ref{zphot} the position of
these galaxies along with their template color type based on the fits of
the SEDs with $LePhare$. We note that the X-ray centroid is within the
contours (see also Fig. \ref{MIPS24} discussed later in the text) of
this concentration of galaxies.

\sethlcolor{red}
\section{Discussion}

\subsection{Identification as a cluster}

\begin{figure}
\centering
\mbox{\epsfig{file=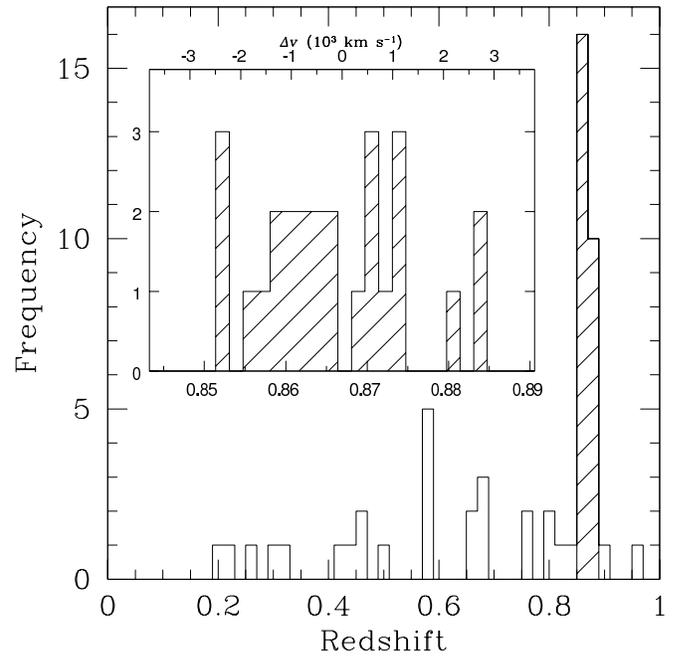,width=3.4in,angle=0}}
\caption{\label{histos} Histogram of the GMOS redshifts, with the region around
0.87 expanded in the insert.}
\end{figure}

The first evidence that RX J1257 is a massive cluster, is the strong
peak at $z \sim 0.87$ seen in Fig.\ref{histos}.

Second, although there is a relatively uniform spatial distribution of
galaxies at the cluster redshift, there is also a significant concentration at
the putative cluster location as defined by the galaxy concentration contours
(see Fig.~\ref{zphot}) and the X-ray emission (see Figs.~\ref{complex} and \ref{color_yellow}).
Thus, this system is likely a cluster embedded in a larger filament or sheet
of galaxies, similar to the sheet of galaxies detected at $\sim 0.735$ in the
CDFS (e.g. Adami et al. 2005).

Third, we applied the SG (Serna $\&$ Gerbal 1996) method to our
redshift catalog. This hierarchical method allows substructures to be
extracted from a catalog containing positions, magnitudes, and
redshifts, based on the calculation of their binding energies. It 
provides a robust velocity dispersion and a virial mass.  The SG method
reveals the existence of 
%
%
a group of 18 galaxies located at the candidate cluster center,
coincident with the X-ray emission, with a mass of $6.1 \times
10^{14}$~M$_{\odot}$, based on a velocity dispersion of 600~km s$^{-1}$.
Within this group, we found two {\it subgroups} of 5 and 3 galaxies with
respective masses of $\sim 10^{14}$~M$_{\odot}$ (velocity dispersion of
289 km s$^{-1}$) and $\sim$2.2 10$^{13}$~M$_{\odot}$ (velocity
dispersion of 255 km s$^{-1}$). These masses from galaxy velocity
dispersion measurements are consistent with the inferred X-ray mass of
$(1-5)\times 10^{14}$ \Ms, and are in the range to call RX J1257 a
massive cluster (e.g. Sarazin 1986).

Fourth, the X-ray emission is clearly extended, as seen in
Figs.~\ref{gastao_color} and \ref{beta} and is similar in size to the
galaxy concentration shown in Fig. \ref{zphot}.

Fifth, assuming a redshift of 0.866, the projected size of the X-ray
emission is about 0.5 Mpc, and its kT (3.6 keV) and inferred L$_{\rm {X,
bol}}$ ($2\times 10^{44}$ \es) are all consistent with a massive
cluster at 0.866 (see, \textit{e.g.}, Ettori et al. 2004a, and
Fig.~\ref{ettori_lx}).

\begin{figure}[h]
\centering
\caption{\label{complex} A 11 arcsecond Gaussian smoothed \cha\ image in
the 0.3--2.0 keV range.  The cyan contours are based on this \cha ~image
and are drawn to highlight the \cha\ data. The red contours are based on
the \xmm\ 0.3$-$6 \keV\ image (see $\S$3.2.) and they coincide with
those in green or yellow in Fig. \ref{color_yellow}. The text in the
figure points to the two brightest regions in the \cha\ image that are
centered in \xmm\ contours. }
\end{figure}

\begin{figure*}[hbt]
\centering
\caption{\label{color_yellow} A color image based on {\it Spitzer } IRAC
3.6 $\mu$m (R), Gemini z' (G) and Gemini i' ( B) image.  The \xmm\ contours
in the 0.3-6 keV interval are shown in green, left, and
yellow/white, right (adjusted to
separate the northern AGN from the cluster emission).  The green circles
show galaxies at the cluster redshift.  The magenta and blue circles indicate
galaxies in the two sub-groups detected by the SG (Serna $\&$ Gerbal 1996)
method; only 3 of the 5 objects in the western sub-group are shown here. 
Big and small white circles indicate the presence or absence of
[OII] emission: no white circles = no detectable [OII], 
small white circle (only two, one in the southeast corner of the image,
the other just  north of the guide star mask) = [OII]
smaller than 2 times the noise, large white circles = [OII] stronger
than 2 times the noise.  The rightmost image is a blow up of the left
one.  North is up and East to the left; a concordant cosmology is
assumed, thus 463 kpc = 1~arcmin (as depicted by the horizontal white
bars in both images) at the redshift of RX J125712+4738.17 ($z=0.866$).  }
\end{figure*}

\subsection{Is RX J1257 bimodal?}
\label{bimodal}

The \xmm\ data suggest the X-ray emission is bimodal, but the \cha\ data show
there is a point source located in the western portion of the X-ray emission.
We show in Fig. \ref{complex} a smoothed (by an 11 arcsec radius Gaussian
kernel) low energy (0.3--2.0 keV) \cha\ image with XMM contours (red)
superposed. There are clearly two centroids enclosed by the \xmm\ contours
(red in Fig.~\ref{complex}) and the west-most centroid has a round (point)
source near the center. This point source is probably the major contribution
to the western peak in the \xmm\ contours associated with the cluster.
Nevertheless, as can be seen in Fig.~\ref{complex} it appears that diffuse
emission surrounds this point source. 
If the point source were absent, the
surrounding diffuse emission could be the origin of the second peak in the
\xmm\ contours. We conclude therefore, that there is still a centroid in the
western X-ray contours that is possibly not due to the point source (the
nature of which is discussed below), and that the X-ray emission associated
with RX J1257 has bimodal characteristics.

This hypothesis of bi-modality fits well with our finding that the
distribution of galaxy cluster members is also bimodal (based on the SG
method) with the two subgroups also aligned in the E-W direction. That the
western subgroup of galaxies lies outside the \xmm\ contours may indicate the
relatively low dynamical age of the cluster (see Fig.~\ref{color_yellow}).
This is confirmed by Fig.~\ref{zphot} where we see that late type galaxies are
relatively numerous in RX J1257, being about half of the cluster population in
the central area.

\subsubsection{Nature of the embedded point source}
\label{pt_src}

\begin{figure}[h]
\centering
\caption{\label{3_in_a_row} $i'$-band image of three blended galaxies
  that run from North to South in the image. The two lower most probable
  positions are marked by white circles. The lowest declination object
  was detected in the \cha\ image with probably diffuse emission around
  it (see text). The next object to the North is most likely a MIPS
  source which is the 7th brightest MIPS object in the field. This
  region can be found in the northernmost of the two magenta circles
  that overlap in Fig \ref{color_yellow}. North is up and East is to the
  left}
\end{figure}

The point source shown in Fig.~\ref{3_in_a_row} and located at
$\alpha$=194.2930, $\delta$=47.6339 (from the \cha\ data) could be associated
with an optical object located at the approximate coordinates
$\alpha$=194.2937, $\delta$=47.6340. However, due to the proximity in
projection of three objects, SExtractor provided only one detection. With this
caveat in mind, we estimate that the probability of this object to be a
cluster member is 8\%\footnote{This was done by integrating the redshift
probability distribution given by LePhare over the total redshift range. We
then integrated over an estimated redshift range for the cluster of 0.8--0.9,
and calculated the ratio of these two values. See Adami et al. (2008).}. The
nature of this object (normal galaxy or active object) remains unclear,
however, because a galaxy template provides a better fit to the {\it
composite} spectrum than a QSO template, but the X-ray flux is $4.8 \times
10^{-13}$ \ecmss\ which at the cluster distance means its L$_{\rm X}$ is about
$10^{45}$\es, typical of QSOs. The most conservative explanation, then, is that
this X-ray source is a chance alignment with the cluster and the derived
photo-z (and spectral classification) for the merged (by SExtractor) source
does not apply to the X-ray source.

\begin{figure}[bht]
\centering
\mbox{\epsfig{file=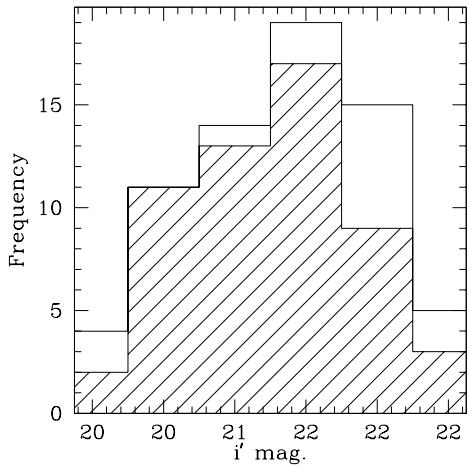,width=2.4in,angle=0}}
\caption{\label{complet} Histogram of the distribution of the magnitudes of the galaxies for which spectra were taken. 
The cross hatched area is for those galaxies for which redshifts were successfully derived.}
\end{figure}

\subsection{Characteristics of the cluster galaxies}

\subsubsection{Selection Effects}
In Fig. \ref{complet} we show the brightness distribution of the galaxies for which we attempted to determine redshifts.
>From this, it can be seen that we were successful a very large fraction of the time and that there was no initial 
choice that
would have favored emission line galaxies. Also, it was possible to determine redshifts even when emission lines
were {\it not} present (over 50\% of the time).
Thus, we conclude that the relatively large faction of emission line galaxies 
we found at the cluster redshift was not due to a selection effect, and we proceed in the following discussion based on
this assumption. 

\subsubsection{Spectra} We first consider those galaxies with spectroscopically determined 
redshifts within the putative cluster redshift range (0.850-0.874).   
We find that about 45\% (10/21) of the cluster galaxies are emission 
line objects (cf. Fig.~\ref{color_yellow}). These results, although based 
on small number statistics, are consistent with the hypothesis that RX J1257 
is relatively young and there has not been time for the cluster environment 
to cause the galaxies to change from late type (star forming) to early type (passively evolving). In comparison,
in more mature low $z$  clusters,  $\sim$75\% of galaxies can be classified 
as early type objects (e.g. Adami et al. 1998).

The fraction of MIPs detections within the population of
spectroscopically confirmed galaxy members is nearly 100\% (19/21).
This high fraction of MIPs detections could be explained if galaxy
harassment has triggered star formation or if the system is so young
that dust has not yet been swept from the late type galaxies.  That the
large majority of the MIPS cluster members in the {\it core} are
associated with the western (rather than eastern) portion of the cluster
X-ray emission (see Fig. \ref {MIPS24}) is suggestive that the western
half is the younger part of the cluster.  The data also suggest that the
system to the west is in the process of merging with the group to the
east.  Therefore, an appealing scenario that explains our results is
that proposed by Elbaz et al. (2007) where late type galaxies at
z$\sim$0.9 are red due to dust.  These galaxies do not become blue until
they have lost enough dust, which could come about through a variety of
evolutionary processes.

\begin{figure*}[bht]
\centering
\caption{\label{MIPS24} Left: a MIPS 24$\mu$m (R) , IRAC-3.6 $\mu$m (G)  and 
i$'$-band (B) image
  that shows all the spectroscopically determined members that were detected
  by the MIPS (red boxes; based on MIPS-i-band source association that is not
  necessarily correct to better than 2 arc seconds) with our observations.
  The circle is 3~arcmin radius or about 1.3 R$_{\rm v}$ (see text). Right: the
  same color scheme on a zoomed scale with the i-band (blue) now greatly suppressed to bring out the MIPs flux. The green contours are based on the
  0.3-2 keV \xmm ~image. North is up and East is left} 
\end{figure*}

\subsection{ Cluster Collapse}
\label{collapse}

\begin{figure}[h]
\centering
\mbox{\epsfig{file=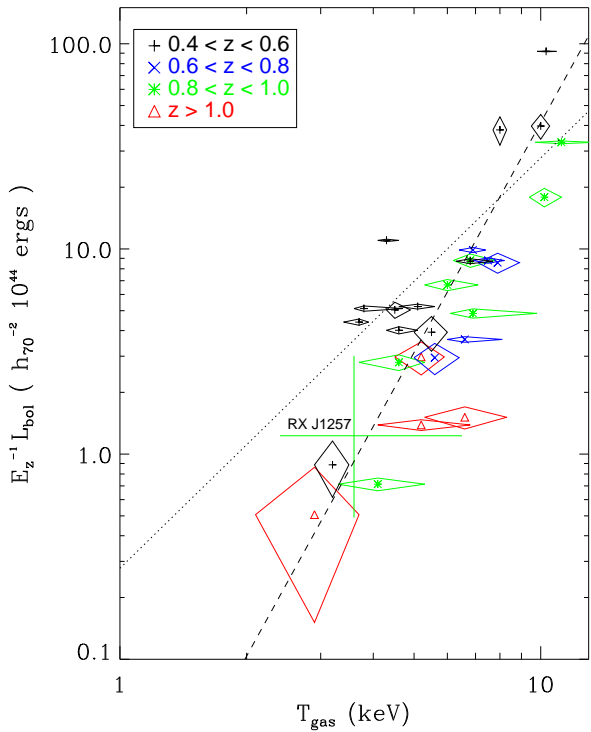,width=3.6in,angle=0}}
\vspace{-.25 in}
\caption{\label{ettori_lx} Based on the Ettori et al. (2004a) Fig. 4, right upper
  panel, with the data point added for RX J1257 as a green cross. This is the L $-$ T relation with
  correction by E($z$).  Dotted line: slope fixed to the predicted value of
  2. 
Dashed line: slope free. See Ettori et al. (2004a) 
for further details.}
\end{figure}

We suggest this cluster is just in the process of formation based on:
(a) the bimodal distribution of both the X-ray emission and the galaxy
population; (b) the fact that the majority of the spectroscopically
confirmed cluster members were detectable in the MIPS 24~$\mu$m channel;
(c) the existence of cluster galaxy substructures based on the
Serna-Gerbal method; (d) the kT on the high side relative to the
predicted L$_{\rm{X,bol}}$-kT relation (the thin {\it dotted} line
in Fig.~\ref{ettori_lx}).

If there is some energy input prior to infall, this would produce a
higher kT versus L$_{\rm{X,bol}}$ than we would expect from 
energy input from the infall alone. For example, the initial energy
injection in addition to gravitational infall comes from some other
process such as galaxy outflows or SN (see Babul et al. 2002, Jones et
al. 2003, Bode et al. 2007). 
Since we have found several late-type cluster member galaxies, this
suggests that galactic outflows have not dominated the heating of the RX
J1257 ICM, otherwise gas and dust loss should have made these galaxies
faint rather than bright 24 $ \mu$m objects.

Another way to consider
Fig.~\ref{ettori_lx} is to assume a lower than normal L$_{\rm{X,bol}}$ for a
given kT. Such a low L$_{\rm{X,bol}}$ is expected by
Ventimiglia et al. (2008), who carried out calculations that showed that clumpy,
not fully merged clusters will tend to have luminosities that fall below
the L$_{\rm{X,bol}}$-kT relation for a given kT.  Thus, the X-ray data
are consistent with our expectations that the cluster is in the process
of merging or collapsing and is not fully relaxed.

\subsection{A Comprehensive Picture}

The results presented here combined with previous work (references
above) on clusters of galaxies lead to a
scenario in which part of the red galaxies in high $z$ clusters
are red due to dust rather than being red and dead.
The less massive the cluster, the younger it will be in terms of having
not yet completed infall, compared to more massive clusters. A system
such as RX J1257 should then, as we found here, have a high fraction of
red dusty galaxies. These galaxies will become blue after dust ejection
and then red again after having consumed all their gas.


\section{Summary and Conclusions}

We have discovered the young massive cluster of galaxies RX J1257.2+4738, 
at $z = 0.866$. The cluster has a
luminosity of about $2 \times 10^{44}$ ergs s$^{-1}$, a kT of about 3.6
keV, and an estimated mass of $6 \times 10^{14}$ \Ms. 

The cluster RX J1257 is in the process of collapsing. At birth, there
was possibly significant energy input to the intracluster medium
besides infall which produced a cluster having an  intracluster medium temperature which 
is not the one expected for its luminosity.  The MIPS 24$\mu$m
detections plus [OII]~3727 emission line galaxies imply a high star
formation rate in several of the cluster member galaxies.

In agreement with Elbaz et al. (2007), we suggest that there should be
an evolutionary sequence from red dusty galaxies in a cluster to blue
galaxies and finally to red-dead galaxies.

\acknowledgement {The authors thank the referee for useful remarks.  We
are grateful to the CFHT and Terapix teams.  We acknowledge financial
support from CNES and PNG, CNRS/INSU, and from the CAPES/COFECUB
French-Brazilian cooperation. GBLN acknowledges support from the CNPq
and FAPESP. MPU acknowledges the support of NASA grants
GO7-8144X//NAS8-03060, GO7-8144X//NAS8-03060, and Agmt.\# 1306461 \/\/
NASA NMO710076.

We thank the groups and many individuals responsible for the successful
launch and operation of \cha, \xmm, and \spi\, and for helping us set up
our observations.  We wish to thank Alberto Noriega-Crespo for
assistance with the MIPS post pipeline data reduction. Last but not
least, we thank Emeric Le Floc'h for useful discussions.}

\end{document}